\documentstyle[PASJadd,psfig]{PASJ95}
\newcommand{\vol}{}
\newcommand{\no}{}
\newcommand{\lett}{}
\newcommand{\spage}{}
\newcommand{\rdate}{1999 November 22}
\newcommand{\adate}{2000 February 14}

\begin{document}

\title{Evolution of Cluster Galaxies in Hierarchical Clustering Universes}

\author{Takashi {\sc Okamoto} and Asao {\sc Habe} \\
{\it Division of Physics, Graduate School of Science, Hokkaido University, Sapporo 060-0810}  \\
{\it E-mail(TO): okamoto@astro1.sci.hokudai.ac.jp}}

\abst{Using cosmological $N$-body simulations of critical (SCDM) and
open ($\Omega = 0.3$, OCDM) cold dark matter models, we investigated
the evolution of cluster galaxies.
Based on our numerical simulation,
we constructed merging history trees of the galaxies.
By following their merging history, we could show that the major merger
fractions of the galaxies in cluster-forming regions is roughly proportional
to $(1+z)^{4.5}$ at low redshifts ($z < 2$), and has a steep peak at
$z \simeq 2.5$ and $z \simeq 3$ in SCDM and OCDM, respectively.
We also show that the cluster galaxies are affected by the tidal interaction
after the clusters are formed.
Because the formation redshift of the cluster in SCDM, $z_{\rm form} = 0.15$,
is much more recent than that of the cluster in OCDM, $z_{\rm form} =  1.6$,
the cluster galaxies in SCDM show more rapid evolution by tidal interactions
from $z = 0.5$ than those in OCDM.}

\kword{galaxies: clusters: general --- galaxies: general ---
galaxies: halos --- galaxies: interactions}

\maketitle
\thispagestyle{headings}

\section
{Introduction}

Rich clusters of galaxies are large laboratories for studying galaxy
evolution (Dressler 1984), and their evolution can be followed with
samples out to $z \sim 1$ (Rosati et al. \ 1998). It is well established
that galaxy populations vary with the density of neighboring
galaxies in clusters of galaxies (Dressler 1980) and depend on the distance
from the clusters' centers (Whitmore et al. \ 1993). The increase in the
fraction of blue, star-forming cluster galaxies with the redshift
(Butcher, Oemler 1978, 1984a, 1984b) has also been well established.
It has been suggested that galaxy-galaxy and galaxy-cluster interactions
play important roles in these effects; especially, a major merger of
galaxies produces elliptical galaxies as merger remnants (Barnes 1989, 1996)
and cumulative tidal interactions induce a morphological transformation of
spiral galaxies to S0 galaxies.

Since it is generally believed that cold dark matter dominates the mass in
the universe, we expect that the formation process of dark matter halos
significantly affects the process of the formation and evolution of galaxies.
In this paper, we discuss our high-resolution
cosmological $N$-body simulations which trace the motion of the dark matter
particles and can resolve the galaxy-sized dark halos within a high-density
environment.
We also consider when the above mentioned interactions (i.e., major mergers and tidal
interactions) act on the evolution of the cluster galaxies during
the formation and evolution of clusters in a hierarchical clustering universe.

We should consider the hydrodynamical processes of the baryonic components
in order to follow the evolution of galaxies. However, hydrodynamical simulations,
e.g., smoothed particle hydrodynamics (SPH) simulations, need much more
CPU time than collisionless simulations. Thus, it is difficult to obtain
a wide dynamical range with such simulations.
Here, we restrict ourselves to follow the evolution of dark matter
halos and to use the galaxy tracing method described by Okamoto and Habe
(1999, hereafter Paper I) to obtain merging history trees of galaxies.

Recently, some authors have studied the evolution of the dark matter halos
of the cluster galaxies in SCDM (Ghigna et al. \ 1998; Paper I).
The epoch  of the formation of a cluster of galaxies is very sensitive to
the value of the cosmological density parameter, $\Omega_0$
(Richstone et al. \ 1992).
Since the evolution of clusters affects the evolution of galactic halos
within the clusters (Paper I),
it is interesting to compare the evolution of cluster galaxies in various
cosmological models with different values of the density parameter.
Here, we examine two cosmological models: one is the critical
universe ($\Omega_0 = 1$), and the other is the open universe
($\Omega_0 = 0.3$). For both models we assume that the mass of the
universe is dominated by cold dark matter (CDM).

The plan of this paper is as follows:
Techniques and parameters of the $N$-body simulations and the method of
creation of merging history trees of galaxies are described
in section 2. Our results are presented in section 3 and discussed
in section 4.

\section
{Simulation}
\subsection
{The Simulation Data Set}
Our simulations followed the evolution of a isolated spheres of
a radius, $R_{\rm sim}$, in both the standard CDM (SCDM) universe
($\Omega = 1$, $h \equiv H_0/100$ ${\rm km \,  s^{-1} Mpc^{-1}} = 0.5$, $\sigma_8 = 0.67$)
and the open CDM (OCDM) universe
($\Omega_0 = 0.3$, $h = 0.7$, $\sigma_8 = 1$).
The normalisations were chosen to approximately match the observed cluster abundance.
We imposed the constraint of the $3\sigma$ peak with
an 8 Mpc Gaussian smoothed density field at the center of
each simulation sphere in order to obtain a rich cluster (Hoffman, Ribak 1991).
The simulations were performed using a parallel tree-code, which was used
in Paper I.
To obtain a sufficient resolution to follow the evolution of the galaxy-sized halos
with a relatively small number of particles, we used the multi-mass $N$-body
code (Navarro et al. \ 1997; Paper I).
These initial conditions of our simulations were made as follows.

First, only long-wavelength components were used for the realization of
the initial perturbation in the simulation sphere using $\sim 10^5$ particles;
we then performed a simulation with these low-resolution particles.
After this procedure, we tagged the particles which are inside a sphere
of radius 3 Mpc centered on the cluster center at $z = 0$.
Next, we went back to the initial time stage, and then divided each tagged
particles into 64 high-resolution particles according to the density
perturbation that is realized by including additional
shorter wavelength components.
As a result, the total number of the particles became $\sim 10^6$.

We then calculated again the dark matter evolution using high- and
low-resolution particles from the new initial condition.
Our analyses were operated only for the high-resolution particles.
The mass of a high-resolution particle was
$m \simeq 5.5 \times 10^8 h^{-1} \MO$,
and its softening length, $\epsilon$, was set to 5 kpc.

The overall parameters and mass of the clusters
in both simulations at $z = 0$ are listed in table 1.

\begin{table*}
\begin{center}
Table~1.\hspace{4pt}Parameters of simulations.\\
\vspace{6pt}
\begin{tabular*}{15cm}{ccccccccc}\hline\hline\\
MODEL&$N_{\rm h}$&$N_{\rm l}$&$\epsilon_{\rm h}$&$\epsilon_{\rm l}%
$&$m_{\rm h}$&$m_{\rm l}$&$R_{\rm sim}$&$M_{\rm cluster}$\\
    & & &[kpc]&[kpc]&$[h^{-1} \MO]$&$[h^{-1} \MO]$%
&[Mpc]&[$h^{-1} \MO$]\\ \hline
SCDM&958592&91911&5&50&$5.4 \times 10^8$&$3.45 \times 10^{10}$&30&$4.65 \times 10^{14}$\\
OCDM&1186240&94396&5&50&$5.5 \times 10^8$&$3.5 \times 10^{10}$&32&$5.53 \times 10^{14}$\\ \hline
\end{tabular*}
\vspace{6pt}
\par\noindent
Subscripts h and l indicate high-resolution and low-resolution particles, respectively.
\end{center}
\end{table*}
\subsection
{Creation of Merging History Trees of Galaxies}

To create merging history trees of galaxies, we have to identify the
galactic dark halos in the sea of dark matter.
The identification of halos in such environments is a critical step
(Bertshinger, Gelb 1991; Summers et al. \ 1995).
The most widely used halo-finding algorithm, called the
friends-of-friends (e.g., Devis et al. \ 1985), is not acceptable,
because it cannot separate substructures inside of large halos.
Since the DENMAX algorithm (Bertshinger \& Gelb 1991) shows good performance,
we used its offspring SKID (Governato et al. \ 1997).

This algorithm groups particles by moving them along the density
gradient to the local density maximum.
The density field and the density gradient are defined everywhere
by smoothing each particle using the SPH-like method with the neighboring
64 particles.
At a given redshift, only particles with local densities greater than
one-third of the virial density at that epoch are moved to the local
density maximum. This threshold roughly corresponds to the local
density at the virial radius. The final step of the process is
to remove all particles that are not gravitationally bound to
their parent halos.
Here, halos which contain more particles than a threshold number,
$n_{\rm th}$, are identified as galactic halos.
Unless we explicitly state, $n_{\rm th} = 30$ is adopted.
This is a large enough number to inhibit the numerical evaporation
of halos (Moore et al. \  1996a).

The method to create the merging history tree is similar to
that mentioned in Paper I; the details are as follows.

We identify galactic halos with a 0.5 Gyr time interval, which
is restricted by the disk space of our computer.
Since this time interval is shorter than the dynamical time-scale
of the clusters and the fading time-scale of evidence of starburst
in galaxies,
it is sufficiently short to construct the merger trees to investigate the effect
of the merging of galaxies and the tidal interactions.

The three most bound particles in each halo are tagged as tracers.
We consider three cases to construct the merger tree of galaxies.

First, if a halo at $t_{i+1}$, where $i$ is the number of a time stage,
has two or more tracers that are contained in the same halo
at $t_i$, then the halo at $t_{i+1}$ is defined as a {\it next halo}
of the halo at $t_i$. In this case, the halo at $t_i$ is a
{\it progenitor} of the halo at $t_{i+1}$.

Next, we consider the case that some halos at $t_{i+1}$ have one
of three tracers of a halo at $t_i$. The halo that has
a tracer which is more bound in the halo at $t_i$ is
chosen as the next halo of the halo at $t_{i+1}$.

Finally, if none of three tracers of a halo at $t_i$
are contained in any halos at the next time stage ($t_i$),
we refer the particle which is the most bound tracer of the halo at
$t_i$ as a {\it stripped tracer}.
We thus call both the halos and stripped tracers {\it galaxies}
throughout this paper.

We construct the merging history trees of galaxies in this way.
A halo which has more than two progenitors at former time stage
is referred as a {\it merger}.
It often happens that satellite galaxies pass through
a central halo. Such cases should not be considered as merging.
Hence, we check that the tracers of galaxies at $t_{i-1}$,
which are contained in a merger at $t_i$ are still in the
same halo at $t_{i+1}$.

In order to estimate the mass of a stellar component of a galaxy,
we assume that the mass of the stellar component is proportional
to the sum of the masses of its all progenitors
(hereafter we call this mass the {\it summed-up-mass}).
Except for the case in which a large fraction of the stellar
component of the galaxies have been stripped during the
halo stripping, this assumption may be valid.
We estimate the summed-up-mass of a galaxy as the sum of the
summed-up-masses of its all progenitors at the previous time stage.
For a newly forming halo which has no progenitors at a previous time stage,
the summed-up-mass of the halo is set to the mass of the halo.
To consider mass increase due to accretion of dark matter
to the halo after its first identification, we replace
the summed-up-mass with the halo mass when the summed-up-mass is smaller
than the halo mass.

\begin{figure}
\psfig{file=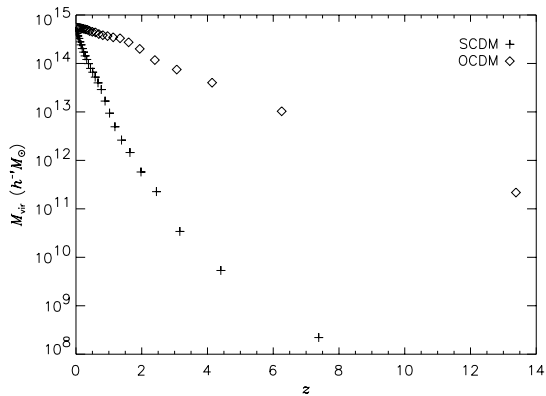,width=8cm}
\caption{Mass evolution of the most massive virialized object
in each model. The plus signs and diamonds represent the mass evolution of
the SCDM cluster and the OCDM cluster, respectively.}
\end{figure}

\section
{Results}

\subsection{Evolution of the Cluster}
To define the size of clusters, we calculated the virial overdensity
based on the spherical-collapse model. For the SCDM
model we used 200 as the virial overdensity according to previous studies
(e.g., Navarro et al. \  1996).
In OCDM, the virial overdensity is a function of redshifts.
Therefore, we calculate it at each redshift and,
it then became $\delta \simeq 400$ at $z = 0$.
In figure 1, we plot the mass evolution of the most massive
virialized object in each model.
It is well known that the formation epoch of the OCDM cluster is much
earlier than that of the SCDM cluster. In our result, the formation
redshift (the redshift when the half of the final mass has accreted)
of the SCDM cluster is $z \simeq 0.15$ and that of the OCDM cluster
is $z \simeq 1.6$. We show the $x$--$y$ projection of a density map in a cube
with sides of $2 \, r_{\rm vir}$ ($r_{\rm vir}$ is the radius of the sphere
having the virial overdensity) centered on the cluster's center
in each model at $z = 0$ (figure 2).  The gray scale represents
logarithmic scaled density given by the SPH like method. We found that
many galaxy-sized density peaks survive even in the central parts
of the rich clusters.

\begin{figure}
\psfig{file=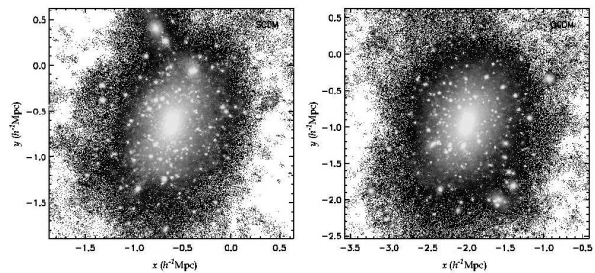,width=8.5cm}
\caption{Density maps of the clusters at $z = 0$. The left panel represents the density
map of the SCDM cluster and right panel represents that of the OCDM cluster.}
\end{figure}

\subsection{Merging of Galaxies}
Numerical simulations have shown that mergers with a mass ratio of 3:1 or
less produce a remnant resembling an elliptical galaxy (Barnes 1996).
Therefore, we define a halo which has more than two ancestors with
this ratio at the former time stage as a ``major merger."
In figure 3, we show the major merger fraction of the
large galaxies
($10^{11} h^{-1} \MO \leq M_{\rm sum} \leq 10^{13} h^{-1} \MO$)
in the cluster-forming regions as a function of the redshift.
The points in the figure can be fitted by a curve in
Gottl\"ober et al. (1999),
$\alpha (1 + z)^{\beta} \exp[\gamma(1+z)]$, with
$\alpha = 0.01, 0.04, \, \beta = 4.5, 4.6,$ and
$\gamma = -1.4, -1.2$ for SCDM and OCDM, respectively.
When the clusters start forming, the merger fraction steeply decreases.
One reason for this is that the high-velocity dispersion
in clusters and groups inhibits the galaxies within these objects
from merging with each other.
Another reason is that the stripping of halos by tidal fields of such large
objects prevents the merging of individual galactic halos
(Funato et al. \ 1993; Bode et al. \ 1994).
Since the cluster in OCDM forms much earlier than in SCDM (see figure 1),
this decline of the merger fraction appears at the higher redshift in
OCDM than SCDM.

\begin{figure}
\psfig{file=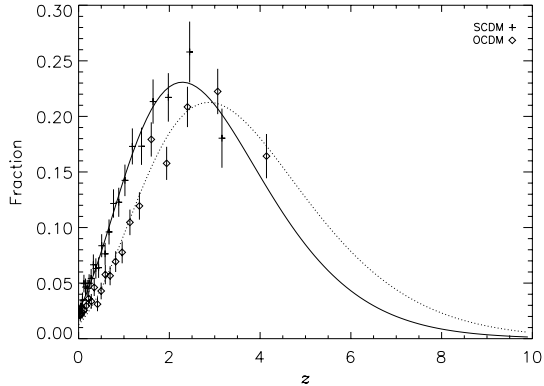,width=9cm}
\caption{Major merger fractions of massive galaxies with
$M_{\rm sum} \geq 10^{11} h^{-1} \MO$ in cluster-forming regions.
The plus signs and diamonds indicate
the fraction in SCDM and in OCDM, respectively.}
\end{figure}

Governato et al. (1999) showed that the major merger rate of the galaxy-sized
dark halos in the field for $z < 1$ is proportional to
$(1 + z)^{4.2}$ and $(1 + z)^{2.5}$ in SCDM and OCDM, respectively.
To compare their result obtained in the field to our result obtained in
the cluster-forming regions, we can say that the efficiency of the major
merging of galaxies in the cluster-forming regions more steeply increases
toward high redshifts than in the field, especially in OCDM.
Since the density contrast of the cluster-forming regions in OCDM
takes a larger value than in SCDM, the evolution of major merger rate in
OCDM is significantly different between in the field and in the
cluster-forming regions.

Recently, van Dokkum et al. (1999) observed the rich cluster at $z = 0.83$,
finding a high merger fraction in the cluster and  rapid evolution of the
fraction. The fraction in $z < 1$ is comparable to the result obtained
here.

\subsection
{Tidal Stripping of Halos}

We evaluated the effects of the tidal stripping on the galactic halos in
different cosmologies.
For this purpose, we chose galactic halos at a redshift when the number of
large halos with $M_{\rm h} \geq 10^{11} h^{-1} \MO$
is largest much before the cluster or group formation epochs.
The tidal effects are probably negligible at such a redshift.
Then, we examined whether the chosen galaxies
loose their halos at lower redshifts.
If they become $M_{\rm h} < 10^{10} h^{-1} \MO$ at lower redshifts, they must be
tidally stripped.
We then adopted $n_{\rm th} = 19, 18$ for SCDM and OCDM, respectively.
Therefore, when a galaxy has become a stripped tracer, it means that
the galaxy does not have a halo with
$M_{\rm h} \geq 10^{10} h^{-1} \MO$.

When they become such small halos,
dissipative effects, which were not included in our simulations,
should become important.
We call such galaxies {\it stripped galaxies}.

In figure 4, we show the stripped galaxy fractions
in the $0.25 h^{-1}$ Mpc radius bins from the cluster centers.
The stripped galaxies show a similar distribution in both models at
$z = 0$. Their evolutions, however, are very different between the models.
In SCDM, there are few stripped galaxies, except for the central
part at $z = 0.5$, and the fraction drastically increases near $z = 0$.
On the other hand, we can see a strong correlation
between the fraction and the radius in the OCDM cluster, even at $z = 0.5$;
this fraction shows very weak evolution during  $z = 0.5$ and $0$.
This is because the cluster in OCDM forms much earlier than in SCDM.

\begin{figure}
\psfig{file=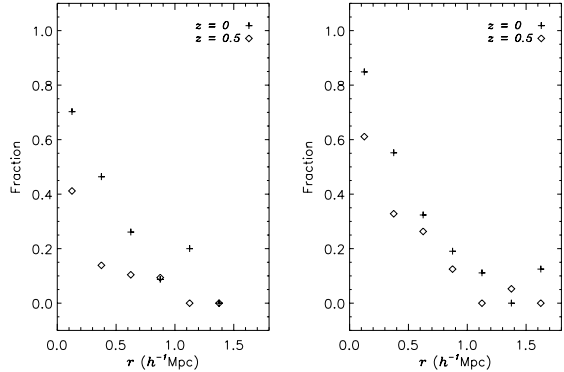,width=9cm}
\caption{Stripped galaxy fractions of galaxies which  have massive halos
($M_{\rm h} \geq 10^{11} h^{-1} \MO$) at the selected redshift in SCDM
(left panel) and in OCDM (right panel).
They are plotted in the $0.25 h^{-1}$ Mpc bins from each cluster center
at $z = 0$ (plus signs) and at $z = 0.5$ (diamonds). }
\end{figure}

Next, we compared the radii of the halos, which were determined
as the radii at which their circular velocity profiles  take minimum values
(Ghigna et al. \ 1998; Paper I), to the tidal radii of the halos estimated
at their pericentric positions, which we calculated by using the dark halo model
of Navarro et al. (1997).
The tidal radii of the halos, $r_{\rm est}$, were estimated by the following
approximation assuming an isothermal distribution of dark matter:
\begin{equation}
r_{\rm est} \simeq r_{\rm peri}\frac{v_{\rm max}}{V_{\rm c}},
\end{equation}
where $v_{\rm max}$ is the maximum value of the circular velocity of a
galactic halo and $V_{\rm c}$ is the circular velocity of a cluster.
In figure 5, we plot $r_{\rm est}$ against $r_{\rm h}$ for the
outgoing halos that must have passed pericenter recently.
We plotted only the large halos ($v_{\rm c} \geq 80$ km ${\rm s}^{-1}$)
to avoid any influence of the insufficient resolution.
Moreover, we ignored the halos with $r_{\rm peri} < 300$ kpc,
because they have tidal tails due to impulsive collisions as
they pass close to the cluster center (Ghigna et al. \ 1998; Paper I).
In SCDM, most of the halos with $r_{\rm h} > 100$ kpc have larger
radii than $r_{\rm est}$.
On the other hand,
the halos in OCDM are sufficiently truncated, and  have comparable
radii to $r_{\rm est}$, because few galaxies  accrete to the cluster
in OCDM at low redshifts
(the cluster has formed at $z_{\rm form} = 1.6$).
This means that the galaxies in the SCDM cluster tidally evolve, even at
present, and such evolution has almost come to completion in the OCDM cluster.
This result is consistent with the weak evolution of the stripped galaxy
fraction of the cluster galaxies in OCDM.

\begin{figure}
\psfig{file=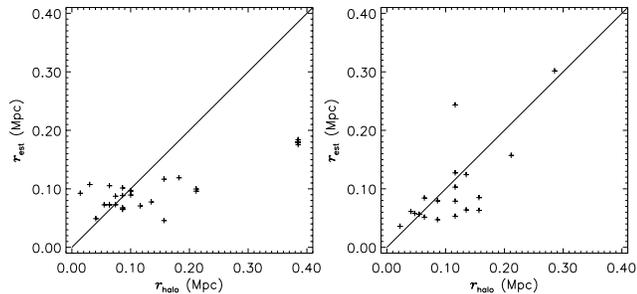,width=9cm}
\caption{Measured values of the radii of the member halos against the expected values,
assuming that the halos have isothermal mass distributions and are
tidally stripped at their pericentric positions.
Only out-going large halos are plotted. The left panel and right panel
represent the halos in the SCDM cluster and those in the OCDM cluster,
respectively.}
\end{figure}

\section{Discussion}

We have investigated the effects of the difference of the cosmological density
parameter on the evolution of the cluster galaxies using the cosmological
$N$-body simulations in SCDM and OCDM.
The cluster forms at $z \simeq 0.15$ and $z \simeq 1.6$ in SCDM and OCDM,
respectively.
We have shown that the difference between the formation epochs of the clusters
changes the evolution of the cluster galaxies.

The major merger fraction in the cluster-forming regions is roughly
proportional to $(1 + z)^{4.5}$ in each cosmological model at low
redshifts ($0 < z < 2$).
The decline of this fraction appears at higher redshifts in OCDM than in SCDM.
The reason is as follows. The efficiency of merging rapidly decreases as
clusters form because of large internal velocities of the clusters and
a reduction of the size of tidally truncated halos. Hence, the earlier
formation of the cluster in OCDM leads to an earlier decline of the major
merger fraction.
From this result, we expect that in a lower density universe the elliptical
galaxies in clusters would mainly form earlier. We will investigate this
possibility in a forthcoming paper.

The tidal interactions also have the possibility to change the morphology
of the galaxies and to induce active star formation (Moore et al. \ 1996b,
1998). We find that, in the SCDM universe, the fraction of cluster galaxies
which have been stripped of their dark halos due to tidal interactions
begins to increase from $z \sim 0.5$. Thus, if the morphological transformation
from S to S0 and the star burst due to the galaxy harassment
(Moore et al. \ 1996b, 1998) are caused by such tidal interactions,
the morphology-density relation and the Butcher--Oemler effect should evolve
from $z \simeq 0.5$ in SCDM.
On the other hand, in OCDM, sice this fraction has already been significant
at $z = 0.5$, we can observe these effects at higher redshifts
than $z = 0.5$.
For detailed analyses we need a star-formation history in each galaxy,
which will be considered in future work.

In this paper we show how the cluster evolution affects on the evolution
of cluster galaxies. The effects of different formation epoches of
the clusters of galaxies between cosmological models on the color
and morphological evolution of the cluster galaxies will be clarified
by combining the merging history trees obtained here with a simple model
of gas cooling, starformation, and feedback used in semianalytical
work (Kauffmann et al. \ 1993; Cole et al. \ 1994; Kauffmann et al. \ 1999).
\par
\vspace{1pc}\par
%
%
We wish to thank M. Fujimoto and M. Nagashima for useful discussions.
Numerical computation in this work was carried on the HP Exemplar
at the Yukawa Institute Computer Facility and on the SGI Origin 2000
at the Division of Physics, Graduate School of Science, Hokkaido University.

\section*{References}

\re
Barnes J.E.\  1989, Nature 338, 123

\re
Barnes J.E. \ 1996, in Formation of the Galactic Halo, Inside and Out,
                ed H.\  Morrison, A.\ Sarajedini
				ASP Conf. Ser. \ 92, p415

\re
Bertshinger E., Gelb J.M.\  1991, Comput. Phys.\ 5, 164

\re
Bode P.W., Berrington R.C., Cohn H.N., Lugger P.M. 1994, ApJ, 433, 479

\re
Butcher H., Oemler A.Jr\ 1978, ApJ 219, 18

\re
Butcher H., Oemler A.Jr\ 1984a, ApJ 285, 426

\re
Butcher H.R., Oemler A.Jr\ 1984b, Nature 310, 31

\re
Cole S., Arag\`{o}n-Salamanca A., Frenk C.S., Navarro J.F., Zepf S.E.\ 1994, MNRAS 271 781

\re
Davis M., Efstathiou G., Frenk C.S., White S.D.M. \ 1985, ApJ 292, 371

\re
Dressler A.\ 1980, ApJ 236, 351

\re
Dressler A. 1984, ARA\&A 22, 185

\re
Funato Y., Makino J., Ebisuzaki T. \ 1993, PASJ, 45, 289

\re
Ghigna S., Moore B., Governato F., Lake G., Quinn T., Sadel J. \ 1998, MNRAS, 300, 146

\re
Governato F., Moore B., Cen R., Stadel J., Lake G., Quinn T.\ 1997, NewAstron. 2, 91

\re
Gottl\"ober S., Klypin A., Kravtsov A.V.\ 1999, astro-ph/9909012

\re
Hoffman Y., Ribak E.\ 1991, ApJ 380, L5

\re
Kauffmann G., Colberg J.M., Deaferio, A., White S.D.M.\ 1999, MNRAS 303, 188

\re
Kauffmann G., White S.D.M., Guiderdoni B.\ 1993, MNRAS 264, 201

\re
Moore B., Katz N., Lake G.\ 1996a, ApJ 457, 455

\re
Moore B., Katz N., Lake G., Dressler A., Oemler A.Jr\
1996b, Nature 379, 613

\re
Moore B., Lake G., Katz N.\ 1998, ApJ 495, 139

\re
Navarro J.F., Frenk C.S., White S.D.M. \ 1997, ApJ 490, 493

\re
Okamoto T., Habe A.\ 1999, ApJ 516, 591 (Paper I)

\re
Richstone D., Loeb A., Turner E.L.\ 1992, ApJ 393, 477

\re
Rosati P., Ceca R.D., Norman C., Giacconni R.\ 1998, ApJ 492, L21

\re
Summers F.J., Davis M., Evrard A.E.\ 1995, ApJ 454, 1

\re
van Dokkum P.G., Franx M., Fabricant D.,  Kelson D.D., Illingworth G.D. \ 1999, ApJ 520, L95

\re
Whitmore B.C., Gilmore D.M., Jones C. \ 1993, ApJ 407, 489

\end{document}